\documentclass[reprint,nofootinbib,superscriptaddress,amsmath,amssymb,aps,prd]{revtex4-2}
\usepackage{amsmath,mathrsfs,amsbsy,color,graphicx,bm,amsthm,amsfonts}
\usepackage{bbm}
\usepackage{times}
\usepackage{units}
\usepackage{dcolumn}
\usepackage{graphicx}
\usepackage{epsfig}
\usepackage{epstopdf}
\usepackage[colorlinks,linkcolor=red,anchorcolor=green,citecolor=blue,CJKbookmarks=True]{hyperref}
\DeclareMathSymbol{\shortminus}{\mathbin}{AMSa}{"39}
\usepackage{mathrsfs}
\usepackage{braket}
\usepackage{amssymb}
\usepackage{txfonts}
\usepackage{float}
\usepackage{enumitem}
\usepackage{multirow}
\usepackage{orcidlink}

\begin{document}
\title{Survival of nonclassical correlations in  Lorentz-violating spacetime}
\author{Yangchun Tang}
\affiliation{Department of Physics, Key Laboratory of Low Dimensional Quantum Structures and Quantum Control of Ministry of Education, Institute of Interdisciplinary Studies,  Hunan Research Center of the Basic Discipline for Quantum Effects and Quantum Technologies,  and Synergetic Innovation Center for Quantum Effects and Applications, Hunan Normal
	University, Changsha, Hunan 410081, P. R. China}

\author{Zhilong Liu~\orcidlink{0009-0000-0353-5113}}
\affiliation{Department of Physics, Key Laboratory of Low Dimensional Quantum Structures and Quantum Control of Ministry of Education, Institute of Interdisciplinary Studies,  Hunan Research Center of the Basic Discipline for Quantum Effects and Quantum Technologies,  and Synergetic Innovation Center for Quantum Effects and Applications, Hunan Normal
	University, Changsha, Hunan 410081, P. R. China}

\author{Wentao Liu\orcidlink{0009-0008-9257-8155}}
\email[]{wentaoliu@hunnu.edu.cn} 
\affiliation{Lanzhou Center for Theoretical Physics, Key Laboratory of Theoretical Physics of Gansu Province, 
	Key Laboratory of Quantum Theory and Applications of MoE,
	Gansu Provincial Research Center for Basic Disciplines of Quantum Physics, 
	Lanzhou University, Lanzhou 730000, China}
\affiliation{Institute of Theoretical Physics $\&$ Research Center of Gravitation,
	Lanzhou University, Lanzhou 730000, China}

\author{Jieci Wang\orcidlink{0000-0001-5072-3096}}
\email{jcwang@hunnu.edu.cn (Corresponding author)}
\affiliation{Department of Physics, Key Laboratory of Low Dimensional Quantum Structures and Quantum Control of Ministry of Education, Institute of Interdisciplinary Studies,  Hunan Research Center of the Basic Discipline for Quantum Effects and Quantum Technologies,  and Synergetic Innovation Center for Quantum Effects and Applications, Hunan Normal
	University, Changsha, Hunan 410081, P. R. China}

\textcolor{blue}{}

\begin{abstract}
The breakdown of Lorentz invariance, a potential signature of quantum gravity, offers a window into physics beyond general relativity. We investigate how such a violation, embodied by the Einstein-Bumblebee black hole spacetime, influences the nonlocal quantum correlations. Specifically, we study the quantum steering and Bell nonlocality between modes trapped inside and outside the event horizon of an Einstein-Bumblebee black hole.  Our analysis demonstrates that quantum steering for an initially correlated state is confined to a narrow region near the event horizon, with the Lorentz-violating parameter further constraining this domain.  Notably, the degree of steering asymmetry is significantly modulated by both the distance from the horizon and the Lorentz-violating parameter, with the two spatially separated regions exhibiting opposite trends. Furthermore, the Bell nonlocality measurable by an external observer strengthens with increasing distance from the black hole. These findings confirm the persistence of nonclassical correlations in a Lorentz-violating gravitational background and  and offer a novel perspective on the interplay between quantum information and fundamental spacetime symmetries.
	\end{abstract}
\vspace*{0.5cm}
\maketitle

\section{Introduction}

In 1935, Einstein, Podolsky, and Rosen (EPR) highlighted the counterintuitive implications of quantum mechanics by arguing that entangled systems exhibit ``spooky" action at a distance, thereby challenging local realism~\cite{Einstein:1935rr}. Shortly afterward, Schrödinger introduced the concept of steering to describe the phenomenon where measurements performed on one subsystem of an entangled bipartite state can remotely influence the state of a distant partner~\cite{Schrodinger:2008pyl}.  The crucial distinction of quantum steering from entanglement resides in the fact that it is inherently asymmetric \cite{Reid:1989zvj,Wiseman:2007hyt,Jones:2007spa}. This directional control allows one party to remotely prepare the state of another through its choice of measurements, regardless of whether the reverse hold~\cite{Handchen:2012enc,Lee:2022kuj,Sohail:2023usg,Guo:2023lkh,Peng:2023xws,Branciard:2012bbh,ElQars:2023gkt,Cao:2018lnk}. 
A more stringent manifestation of non-classicality is Bell nonlocality, which arises when correlations violate Bell inequalities and cannot be explained by any Local Hidden Variable ( theory~\cite{Goyeneche:2020xtv,Brunner:2013est,Broadbent:2006zgz}. These concepts form a strict hierarchy within quantum correlations in flat spacetime: all Bell-nonlocal states are steerable, and all steerable states are entangled, but the converse implications do not generally hold~\cite{Uola:2020kps,Chen:2016iic,Chen:2018oxm}. 

On the other hand, the rapid development of quantum information science has spurred intense interest in the resilience of these correlations in relativistic settings. A growing body of work demonstrates that entanglement, steering, and Bell nonlocality are highly sensitive to external conditions such as spacetime curvature, acceleration, and the presence of event horizons
~\cite{Fuentes-Schuller:2004iaz,Wu:2023spa,Liu:2025hcx,Liu:2025bzv,Tang:2025mtc,Liu:2024yrf,Cong:2018vqx,Zhang:2020xvo,Gallock-Yoshimura:2021yok,Zhou:2021nyv,Bueley:2022ple,Chatterjee:2025pky,Lopez-Raven:2025ehf,Wang:2025lga}. Given their extreme gravitational environments, black holes serve as a premier theoretical laboratory for probing the resilience of these quantum resources. Furthermore, while Lorentz invariance (LI) remains a foundational symmetry of relativity, several candidate theories of quantum gravity suggest its violation or spontaneous breaking due to the presence of new fundamental fields or high-energy corrections to the underlying spacetime structure~\cite{Mattingly:2005re,Amelino-Camelia:2008aez}. This spontaneous breaking effect has prompted extensive research into modified gravity frameworks~\cite{Liu:2024wpa,Liu:2025bpp,Tang:2025eew,Kostelecky:2003fs,Du:2025lkp,Xiao:2025flt,Kostelecky:2025zsy,Tian:2025wfu,Tian:2022gfa,AraujoFilho:2025nmc}. A representative model is the Einstein-Bumblebee framework~\cite{Kostelecky:2003fs,Bluhm:2007gs}, in which a vector field with a nonminimal coupling to curvature acquires a nonzero vacuum expectation value, thereby inducing spontaneous Lorentz violation~\cite{Casana2018,Ovgun2019,Gullu2020,Liu:2022xse,Liu:2022dcn,Maluf2021,Xu:2022frb,Ding2022,Poulis:2021nqh,Zhang:2023wwk,Lin:2023foj,Chen2020,Mai:2024lgk,Hosseinifar:2024wwe,Finke:2024ada,Liu:2024axg,Li:2025itp,Liu:2025oho,Lai:2025nyo,Chen:2025ypx,Sekhmani:2025zen,Shi:2025tvu}. 

In this work, we investigate the interplay between spontaneous Lorentz symmetry breaking of spacetime and quantum correlations, specifically quantum steering and Bell nonlocality, in the spacetime of an Einstein-Bumblebee black hole. We quantify how these non-classical correlations, shared between observers in physically accessible and inaccessible regions, are modulated by both the distance from the event horizon and the fundamental Lorentz-violating parameter of the Bumblebee model. Our analysis reveals distinct, asymmetric behaviors in steering and a persistent signature of Bell nonlocality, offering new insights into the survival of quantum information in regimes where standard spacetime symmetries may break down.

The paper is organized as follows. In Sec. \ref{sec2}, we quantize the Dirac field within the Einstein-Bumblebee black hole background. Sec. \ref{sec3} presents the theoretical framework for quantum steering. Sec. \ref{sec4} investigates the induced nonlocal behavior of fermionic fields under Lorentz violation, and Sec. \ref{sec5} explores the influence of black hole gravity on Bell nonlocality. Finally, Sec. \ref{sec6} provides a summary and the conclusions.

\section{Quantization of the Dirac Field in the Context of Black Holes} \label{sec2}

The metric expression for a static Einstein-Bumblebee black hole is given by~\cite{Casana2018} \begin{equation}
	\label{eq:metric}
	ds^{2} = -\left(1 - \frac{2M}{r}\right) dt^{2} + \frac{1 + \ell}{1 - \frac{2M}{r}} dr^{2} + r^{2} d\Omega^{2},
\end{equation}
where $d\Omega^{2}$ denotes the angular portion of a unit sphere, $\ell$ denotes the Lorentz-violating parameter, $r$ denotes the distance from the observer at a fixed position to the event horizon of the black hole, and $M$ denotes the mass of the black hole.
Defining $H(r) = 1 - \frac{2M}{r}$, the event horizon is determined by the condition $H(r) = 0$, which yields $r_{h} = 2M$. Considering only the radial coordinate and adopting the near-horizon approximation, we obtain
\begin{equation}
	\label{eq:metric1}
	ds^{2} = -(r - r_{h}) H^{\prime}(r_{h}) dt^{2} + (1 + \ell) (r - r_{h})^{-1} H^{\prime}(r_{h}) dr^{2}.
\end{equation}
By substituting the coordinate transformation $dr = \frac{\zeta H'(r_h)}{2(1+\ell)} d\zeta$ into the equation (\ref{eq:metric}), the metric  is  found to be
\begin{equation}
	\label{eq:metric2}
	\begin{split}
		& ds^{2} = -\left[ \frac{c_{1} - r_{h}}{H^{\prime}(r_{h})^{-1}} + \frac{\zeta^{2} H^{\prime}(r_{h})^{2}}{4(1 + \ell)} \right] dt^{2} \\
		& \qquad + \left[ 1 + \frac{4(c_{1} - r_{h})(1 + \ell)}{\zeta^{2} H^{\prime}(r_{h})} \right]^{-1} d\zeta^{2},
	\end{split}
\end{equation}
where $c_1$ denotes an integration constant.
Assuming $c_1 = r_h$ and considering surface gravity $\kappa = \frac{1}{4M(1+\ell)}$ \cite{Wald:1993nt}. 
For an observer located at $r = r_{0}$, introducing  $d\tau = \sqrt{H_{0}} dt$, with $H_{0} \equiv H(r_{0})$, the metric can be rewritten as
\begin{equation}
	\label{eq:metric3}
	ds^{2} = -\frac{(1 + \ell) \kappa^{2} \zeta^{2}}{H_{0}} d\tau^{2} + d\zeta^{2},
\end{equation}
from which we can obtain the acceleration parameter in the  spacetime as $\frac{\sqrt{1+\ell}\,\kappa}{\sqrt{H_0}}$. 

To clarify the  acceleration-induced effect modifies  quantum correlations, it is necessary to evaluate the acceleration at the event horizon of the black hole for a fixed radial position  $r = r_0$.
The four-velocity of a static particle is given by $u^\mu = \{u^0, 0, 0, 0\}$, where $u^0$ is determined by a normalized relation, and $u^\mu u_\mu = -1$. The corresponding acceleration expression is defined as $a^\nu = u^\mu \nabla_\mu u^\nu = u^\mu \partial_\mu u^\nu + \Gamma^\nu_{\mu\rho} u^\mu u^\rho$.
The proper acceleration at a fixed radial position $r$ is given by
\begin{equation}
	a(r) = \sqrt{a^{\mu} a^{\nu} g_{\mu \nu}} = \frac{H'(r)}{2 \sqrt{(1+\ell) H(r)}},
\end{equation}
where $a^{\mu} = \left\{0,\ \frac{H'(r)}{2(1+\ell)},\ 0,\ 0\right\}$ and
$a_{\mu} = \left\{0,\ \frac{H'(r)}{2H(r)},\ 0,\ 0\right\}$.
Setting $r = r_0$, which may be regarded as an approximate radial distance from the event horizon, one obtains
\begin{equation}
	\label{eq:Hprime}
	H'(r) \simeq \frac{\partial}{\partial r}\left[(r - r_h) H'(r_h)\right] = H'(r_h) = 2(1 + \ell) \kappa.
\end{equation}
Substituting this result into the equation (\ref{eq:metric3}), we find
\begin{equation}
	\label{eq:Hprime11}
	ds^{2} = - (a_0 \zeta)^{2} d\tau^{2} + d\zeta^{2}.
\end{equation}
Thus, the explicit expression for the  acceleration at the radial position $r_0$ is found to be 
\begin{equation}
	\label{eq:Hprime22}
	a_0 \equiv a(r_0) = \frac{\sqrt{1 + \ell} \ \kappa}{\sqrt{H_0}},
\end{equation}
which is consistent with the parametric form of the acceleration in Rindler spacetime.
Since $r = r_{h}$ is a coordinate singularity, we consider Kruskal like coordinates and introduce $r_{*} = \sqrt{1 + \ell} \int H(r)^{-1}  dr$ to facilitate extending spacetime beyond them.
Thus, the metric near the black hole becomes \cite{Liu:2024wpa}
\begin{equation}
	ds^{2} = -H(r) e^{-2 \sqrt{(1+\ell) \kappa} r_{*}} \, d\mathrm{U} \, d\mathrm{V} \simeq -e^{-1} d\mathrm{U} d\mathrm{V}.
	\label{eq:metric_approx}
\end{equation}
By considering Kruskal like coordinates,
the differential forms become $du=e^{\sqrt{(1+\ell)k}\,u}dU$, $dv=e^{-\sqrt{(1+\ell)k}\,v}dV$.
Where the null coordinates are defined by the retarded coordinate $u = t - r_{*}$   and the advanced coordinate $v = t + r_{*}$.
According to reference \cite{Martin-Martinez:2010yva}, we can similarly define the physical time-like vectors for these three regions
\begin{align}
	\partial_{t} &\propto \partial_{u} + \partial_{v}, \label{eq:time_deriv1} \\
	\partial_{t} &\propto (u \partial_{u} + v \partial_{v}), \label{eq:time_deriv2}
\end{align}
the expression for $-\partial_t$ is  similar. 
By analyzing different physical time-like vectors, three different types of vacuum states are defined.  
Basing on the quantization of Dirac field, the creation and annihilation operators defined in the static Einstein–Bumblebee black hole spacetime and in Kruskal coordinates are related through Bogoliubov transformations. 
 Then the vacuum and excited states in the Lorentz-violating spacetime from an accelerated observer are found to be \cite{Alsing:2006cj} 
 \begin{align}
	|0\rangle_H &= \cos\alpha\,|0\rangle_R|0\rangle_{\bar{R}} + \sin\alpha\,|1\rangle_R|1\rangle_{\bar{R}} ,\label{eq:state0} \\
	|1\rangle_H &= |1\rangle_R|0\rangle_{\bar{R}} ,\label{eq:state1}
\end{align}
Considering the mode frequency and the ratio between an observer fixed at a certain position and the black hole radius $R_s$, the dimensionless parameters are defined as
\begin{equation}
	\omega = \omega_i M,  R_0 = \frac{r_0}{R_s}.
	\label{eq:alpha_formula11}
\end{equation}
 By substituting the Lorentz-violating parameter, we obtain
\begin{equation}
	\tan\alpha = \exp\left( -\frac{\pi\omega_{i}}{a_{0}} \right) = \exp\left[ -4\pi\omega\sqrt{1+\ell-\frac{1+\ell}{R_{0}}} \right].
	\label{eq:alpha_formula}
\end{equation}
This equation provides an exact expression and can be employed to analyze how the acceleration parameter and the Lorentz-violating parameter vary with distance from the event horizon.

\section{Quantification of two-part steering} \label{sec3}
For a bipartite quantum system,  the general form of the density matrix $\rho_x$  describing the bipartite state is 
\begin{equation}
	\rho_x = \begin{bmatrix}
		\rho_{11} & 0 & 0 & \rho_{14} \\
		0 & \rho_{22} & \rho_{23} & 0 \\
		0 & \rho_{23} & \rho_{33} & 0 \\
		\rho_{14} & 0 & 0 & \rho_{44}
	\end{bmatrix},
	\label{eq:rho_x_matrix}
\end{equation}
where  $\rho_{ij}$ ($i,j=1,2,3,4$) represents the element of the density matrix, and the specific element expressions vary depending on the entangled state  \cite{Wootters:1997id}.
Based on the relationships among elements in the two-body system, we describe the local correlations between the states of the two particles. According to the above equations (\ref{eq:rho_x_matrix}), one obtains the specific definition for the concurrence of state X as \cite{Wootters:1997id,Coffman:1999jd}
\begin{equation}
	C(\rho_x) = 2\max\left\{0, |\rho_{14}| - \sqrt{\rho_{22}\rho_{33}}, |\rho_{23}| - \sqrt{\rho_{11}\rho_{44}}\right\}.
	\label{eq:concurrence}
\end{equation}

For the quantum state $\rho_{AB}$ shared by Alice and Rob, if the density matrix $\tau_{AR} = \frac{\rho_{AR}}{\sqrt{3}} + \frac {3 - \sqrt{3}} {3} (\rho_{A} \otimes \frac{I}{2})$ is entangled \cite{Das:2019uhc}, the steering from Rob to Alice is verified. Here, $\rho_{A}$ is the reduced density matrix, expressed as $\rho_{A} = \operatorname{Tr}_{R} (\rho_{AR})$, and $I$ is the two-dimensional identity matrix.
Similarly, if the density matrix $\tau_{RA} = \frac{\rho_{AR}}{\sqrt{3}} + \frac{3 - \sqrt{3}}{3} (\frac{I}{2} \otimes \rho_{R})$ is entangled, then we can determine the steering from Alice to Rob, where the reduced density matrix is $\rho_{R} = \operatorname{Tr}_{A} (\rho_{AR})$.
For the $\rho_x$ matrix in state X,  we employ the definition   \cite{Yu:2007bwc}
\begin{equation}
	\tau_{AR}^{x}=\begin{pmatrix}
		\frac{\sqrt{3}}{3}\rho_{11}+b & 0 & 0 & \frac{\sqrt{3}}{3}\rho_{14} \\
		0 & \frac{\sqrt{3}}{3}\rho_{22}+b & \frac{\sqrt{3}}{3}\rho_{23} & 0 \\
		0 & \frac{\sqrt{3}}{3}\rho_{32} & \frac{\sqrt{3}}{3}\rho_{33}+s & 0 \\
		\frac{\sqrt{3}}{3}\rho_{41} & 0 & 0 & \frac{\sqrt{3}}{3}\rho_{44}+s
	\end{pmatrix},
	\label{eq:tau_AB_x}
\end{equation}
where $b=\frac{3-\sqrt{3}}{6}(\rho_{11}+\rho_{22})$ and $s=\frac{3-\sqrt{3}}{6}(\rho_{33}+\rho_{44})$.

Based on the concurrency equation (\ref{eq:concurrence}) , we know that as long as the inequalities $|\rho_{14}|^{2}>G_{a}-G_{b}$ or  $|\rho_{23}|^{2}>G_{c}-G_{b}$ are satisfied, the matrix $\tau_{AR}^{x}$ can be considered entangled \cite{Horodecki:2000oqp}, where
\begin{align}
	\begin{aligned}
		G_{a} &= \frac{2-\sqrt{3}}{2}\rho_{11}\rho_{44}+\frac{2+\sqrt{3}}{2}\rho_{22}\rho_{33} \\
		&\quad +\frac{1}{4}(\rho_{11}+\rho_{44})(\rho_{22}+\rho_{33}), \label{eq:Ga} 
	\end{aligned}\\
	\begin{aligned}
		G_{b} &= \frac{1}{4}(\rho_{11}-\rho_{44})(\rho_{22}-\rho_{33}) ,\label{eq:Gb} 
	\end{aligned}\\
	\begin{aligned}
		G_{c} &= \frac{2+\sqrt{3}}{2}\rho_{11}\rho_{44}+\frac{2-\sqrt{3}}{2}\rho_{22}\rho_{33} \\
		&\quad +\frac{1}{4}(\rho_{11}+\rho_{44})(\rho_{22}+\rho_{33}). \label{eq:Gc}
	\end{aligned}
\end{align}
Thus, we can verify the steering ability from Rob to Alice. Similarly, the steerability from Alice to Rob can also be verified by satisfying the inequalities $|\rho_{14}|^{2}>G_{a}+G_{b}$ or $|\rho_{23}|^{2}>G_{c}+G_{b}$, proving that $\tau_{AR}^{x}$  is entangled.

The controllability of the steering process from Rob to Alice and from Alice to Rob is  quantified by introducing the variables $S^{R \rightarrow A}$ and $S^{A \rightarrow R}$. Based on the inequality relationship, we can derive the equation for the steerability from Rob to Alice as
\begin{equation}
	S^{R \!\rightarrow A} \!=\! 
	\max \left\{ 0, \frac{8}{\sqrt{3}} \left[ |\rho_{14}|^{2} \!-\! G_{a} \!+\! G_{b} \right], \frac{8}{\sqrt{3}} \left[ |\rho_{23}|^{2} \!-\! G_{c} \!+\! G_{b} \right] \right\},
	\label{eq:steering_B_to_A}
\end{equation}
and similarly obtain the equation for the steerability from Alice to Rob as
\begin{equation}
	S^{A \!\rightarrow R} \!=\! 
	\max \left\{ 0, \frac{8}{\sqrt{3}} \left[ |\rho_{14}|^{2} \!-\! G_{a} \!-\! G_{b} \right], \frac{8}{\sqrt{3}} \left[ |\rho_{23}|^{2} \!-\! G_{c} \!-\! G_{b} \right] \right\}.
	\label{eq:steering_A_to_B}
\end{equation}
By defining the coefficients of the steering relations as $\frac{8}{\sqrt{3}}$, the steering measure of the maximally entangled state attains its maximal value of unity.

\section{Quantum steering in the Context of Lorentzian Defects} \label{sec4}
We consider  Alice and Rob initially share a maximally entangled state in the asymptotically flat region of the Bumblebee black hole spacetime
\begin{equation}
	|\phi_{AR}\rangle = \frac{1}{\sqrt{2}} (|0\rangle_A |0\rangle_R + |1\rangle_A |1\rangle_R),
	\label{eq:phi_AR}
\end{equation}
where modes $A$ and $R$ correspond to Alice's and Rob's respective observations. Owing to Alice remains in the asymptotically flat region, while Rob hovers near the event horizon of the bumblebee black hole. We find that the initial state becomes a three-body system in Einstein-Bumblebee black hole  spacetime \cite{Fuentes-Schuller:2004iaz,Wang:2009qg}
\begin{equation}
	\begin{split}
		|\phi_{AR\bar{R}}\rangle = \frac{1}{\sqrt{2}} \bigl( & \cos \alpha |0\rangle_A |0\rangle_R |0\rangle_{\bar{R}} + \sin \alpha |0\rangle_A |1\rangle_R |1\rangle_{\bar{R}} \\
		& + |1\rangle_A |1\rangle_R |0\rangle_{\bar{R}} \bigr).
	\end{split}
	\label{eq:phi_ARR}
\end{equation}
Here, $A$ represents the mode of subsystem Alice, $R$ represents the mode observed by observer Rob in Einstein-Bumblebee black hole  spacetime, and $\bar{R}$ represents the mode of the subsystem observed by the virtual observer Anti-Rob in Einstein-Bumblebee black hole  spacetime.
Therefore, we obtain the density matrix in the form 
\[
\rho_{AR\bar{R}} = 
\begin{pmatrix}
	\frac{1}{2}\cos^{2}\alpha & 0 & 0 & \frac{1}{2}\sin\alpha\cos\alpha & 0 & 0 & 0 & \frac{1}{2}\cos\alpha \\
	0 & 0 & 0 & 0 & 0 & 0 & 0 & 0 \\
	0 & 0 & 0 & 0 & 0 & 0 & 0 & 0 \\
	0 & 0 & 0 & 0 & 0 & 0 & 0 & 0 \\
	\frac{1}{2}\sin\alpha\cos\alpha & 0 & 0 & \frac{1}{2}\sin^{2}\alpha & 0 & 0 & 0 & \frac{1}{2}\sin\alpha \\
	0 & 0 & 0 & 0 & 0 & 0 & 0 & 0 \\
	0 & 0 & 0 & 0 & 0 & 0 & 0 & 0 \\
	\frac{1}{2}\cos\alpha & 0 & 0 & \frac{1}{2}\sin\alpha & 0 & 0 & 0 & \frac{1}{2}
\end{pmatrix}.
\]
To verify the steering properties of the bipartite quantum states, we analyze the underlying tripartite system by reducing it to three distinct bipartite configurations. 

First, by tracing over the modes associated with anti-Rob, we obtain the reduced mixed density matrix describing the Alice–Rob subsystem, which is given by
\begin{equation}
	\hat{\rho}_{AR} = 
	\begin{pmatrix}
		\frac{1}{2}\cos^{2}\alpha & 0 & 0 & \frac{1}{2}\cos\alpha \\
		0 & \frac{1}{2}\sin^{2}\alpha & 0 & 0 \\
		0 & 0 & 0 & 0 \\
		\frac{1}{2}\cos\alpha & 0 & 0 & \frac{1}{2}
	\end{pmatrix}.
	\label{eq:density_matrix}
\end{equation}
Using the above density matrix, and according to the above equations (\ref{eq:steering_B_to_A}, \ref{eq:steering_A_to_B}), the analytical solution for the quantum steering  between Alice and Rob are found to be
\begin{equation}
	S^{R \rightarrow A} = 
	\text{max}\left[0, \frac{2|\cos\alpha|^{2} + (\sqrt{3}-2){\cos}^2\alpha - {\sin^2}\alpha}{\sqrt{3}}\right],
\end{equation}

\begin{equation}
	S^{A \rightarrow R} = 
	\text{max}\left[0, \frac{2|\cos \alpha|^{2} + \cos^2\alpha~ (\sqrt{3}-2 -\! \sin^2\alpha)}{\sqrt{3}}\right].
\end{equation}

\begin{figure}[h]
	\centering
	\includegraphics[width=0.65\linewidth]{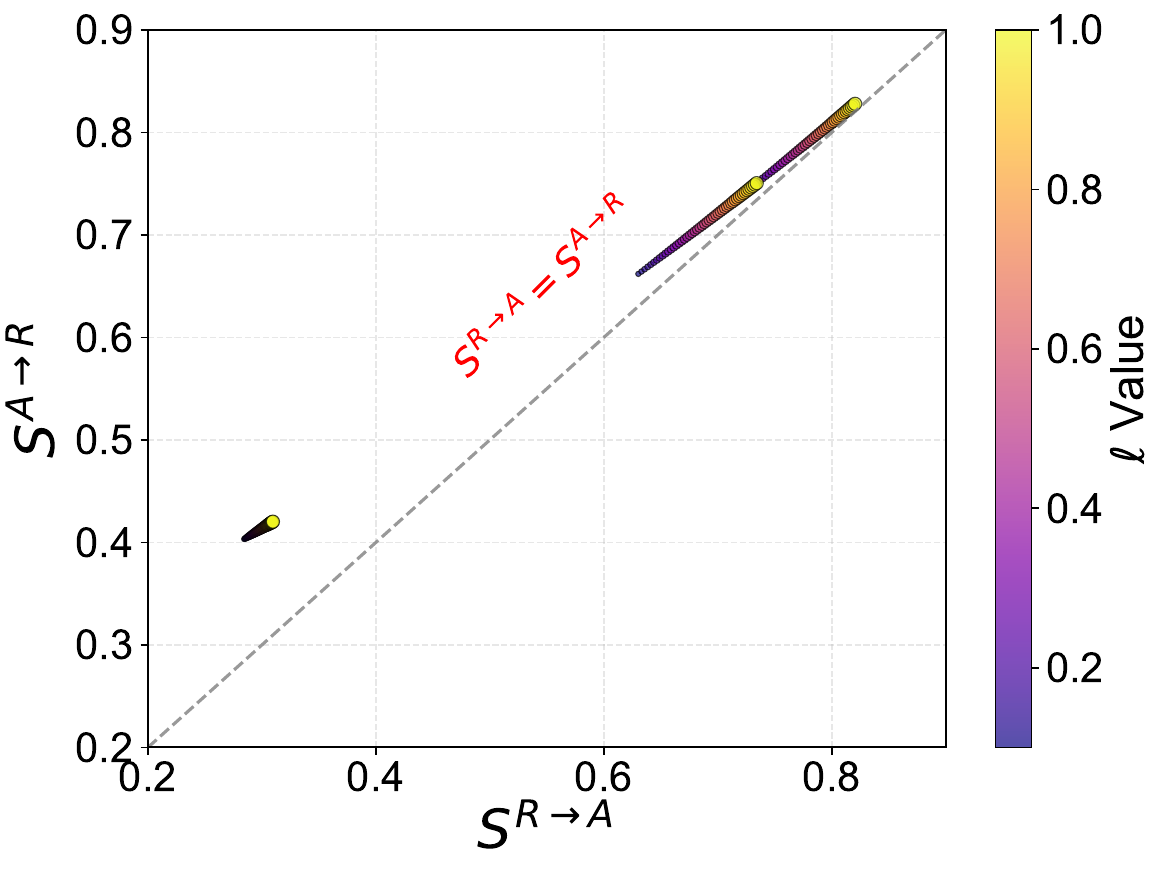}
	\caption{The steering $S^{R \rightarrow A}$ from Rob to Alice and the steering  $S^{A \rightarrow R}$ from Alice to Rob vary with $R_{0}$ and $\ell$. The three segments of $R_{0}$ values from left to right increase from 2 to 3, where the color bar $\ell$ represents values ranging from 0.1 to 1, with $\omega=0.1$.}
	\label{fig1}
\end{figure}

From Fig. \ref{fig1},  we can see that at sufficiently high $\omega$, the steerability between Alice and Rob becomes negligible. 
In contrast, at sufficiently low $\omega$, Lorentz violation effects significantly influence the steerability from Rob to Alice. 
It is shown the horizontal axis represents quantum steering from Rob to Alice, whereas the vertical axis corresponds to quantum steering from Alice to Rob. 
Quantum steering in a bipartite system can be classified into three categories: bidirectional steering, unidirectional steering, and no steering. 
Here, $S > 0$ indicates the presence of quantum steering, whereas $S = 0$ signifies its absence. Interestingly, we find that the steering measure $S^{A \rightarrow R}$ consistently exceeds $S^{R \rightarrow A}$, while the system remains in the bidirectional-steering regime. 

The symmetric relation between $S^{A \rightarrow R}$ and $S^{R \rightarrow A}$ is represented by a dashed line, illustrating the intrinsic symmetry of quantum steerability.
Here, we present three segments illustrating the dependence of the quantum steering measures $S^{A \rightarrow R}$ and $S^{R \rightarrow A}$ on the dimensionless parameter $R_{0}$. 
As $R_{0}$ increases, the quantum steering between Alice and Rob increases. 
On the right side of the figure, a color bar illustrates the effect of varying $\ell$ from 0.1 to 1.0 on the quantum steering between Alice and Rob.
It is worth noting that, for fixed $R_{0}$, the steerability from Rob to Alice increases with increasing $\ell$, and a similar enhancement is observed from Alice to Rob. 
This confirms that even at fixed $R_{0}$, different values of $\ell$ significantly affect the steering behavior between Alice and Rob. 
Thus, we conclude that quantum steering is restricted to a narrow region near the event horizon, and the inclusion of the Lorentz-violating parameter further confines this region.

Next, by tracing over Rob's local modes, the reduced density matrix between Alice and Anti-Rob is  
\begin{equation}
	\hat{\rho}_{A\bar{R}} = 
	\begin{pmatrix}
		\frac{1}{2}\cos^{2}\alpha & 0 & 0 & 0 \\
		0 & \frac{1}{2}\sin^{2}\alpha & \frac{1}{2}\sin\alpha & 0 \\
		0 & \frac{1}{2}\sin\alpha & \frac{1}{2} & 0 \\
		0 & 0 & 0 & 0
	\end{pmatrix}.
	\label{eq:rho_AbarR}
\end{equation}
According to (\ref{eq:steering_B_to_A}, \ref{eq:steering_A_to_B}), the analytical solutions for quantum steering related to the maximally entangled state between Alice and Anti-Rob are
\begin{equation}
	S^{\bar{R} \to A} = 
	\text{max}\left[0, \frac{2|\sin \alpha|^{2} - \cos^2 \alpha + ( \sqrt{3}-2) \sin^2\alpha}{\sqrt{3}} \right],
\end{equation}
\begin{equation}
	S^{A \to \bar{R}} =
	\text{max}\left[0, \frac{2\sin^2 \alpha + ( \sqrt{3} -2- \cos^2 \alpha) \sin^2 \alpha}{\sqrt{3}} \right].
\end{equation}

\begin{figure}[h]
	\centering
	\includegraphics[width=0.48\linewidth]{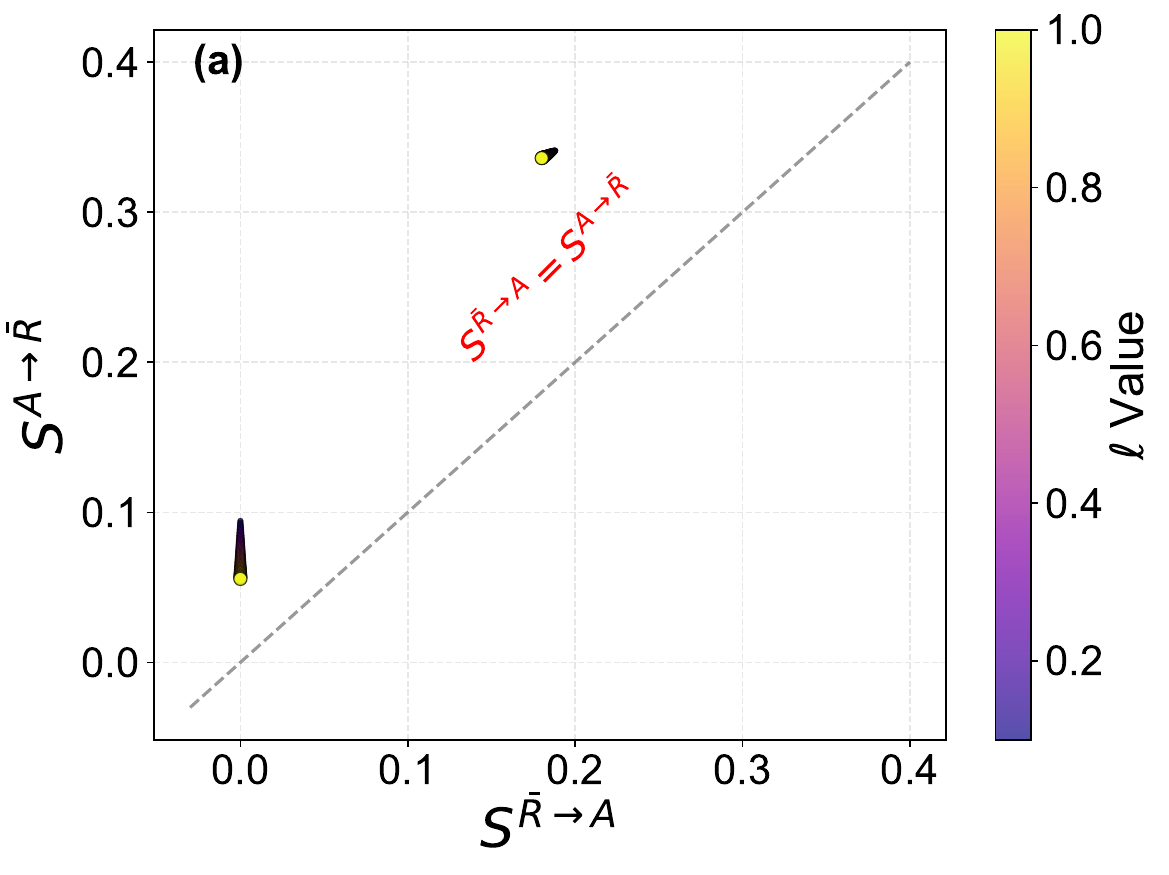}
	\hfill
	\includegraphics[width=0.48\linewidth]{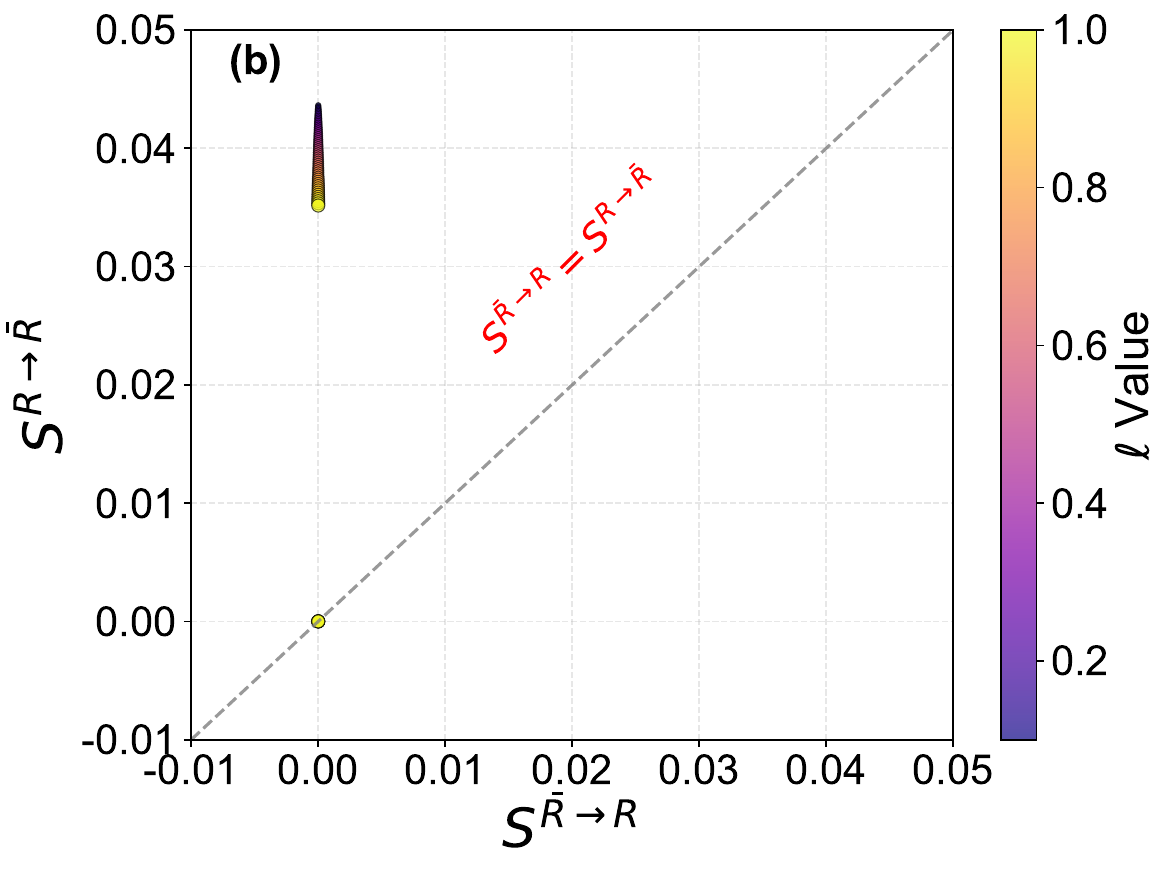}
	\caption{(a) shows the variation of the steering $S^{\bar{R} \to A}$ from Anti-Rob to Alice and the steering $S^{A \to \bar{R}}$ from Alice to Anti-Rob with respect to $R_{0}$ and $\ell$. The values of $R_{0}$ in the two segments range from 2 to 3, with larger $R_{0}$ values on the left and smaller values on the right.
		(b) shows the variation of the steering from Anti-Rob to Rob $S^{\bar{R} \to R}$ and from Rob to Anti-Rob $S^{R \to \bar{R}}$ with respect to $R_{0}$ and $\ell$. The two segments of $R_{0}$ range from 2 to 3, with the upper segment having larger values and the lower segment having smaller values. Here, $\omega=0.1$, and the color bar $\ell$ represents values from 0.1 to 1.}   
	\label{fig2}
\end{figure}

We obtain the steering diagram for Alice to Anti-Rob and Anti-Rob to Alice, as shown in Fig. \ref{fig2} (a). 
It is readily apparent from the figure that, at sufficiently low values of $\omega$, the horizontal axis represents steering from Anti-Rob to Alice, whereas the vertical axis corresponds to steering from Alice to Anti-Rob.  The symmetry between $S^{A \to \bar{R}}$ and $S^{\bar{R} \to A}$ is indicated by a dashed line, interestingly, from which we observe that $S^{A \to \bar{R}}$ consistently exceeds $S^{\bar{R} \to A}$. Simultaneously, color bars corresponding to $\ell$ values ranging from 0.1 to 1.0 show that increasing $\ell$ modifies the steerability between Alice and Anti-Rob. For fixed $R_{0}$, the steerability from Alice to Anti-Rob decreases as $\ell$ increases, and the steerability from Anti-Rob to Alice also decreases with increasing $\ell$. This demonstrates that variations in $\ell$ within the exterior region influence the asymmetry of quantum steering. We also find that when $R_{0}$ is small (i.e., close to the event horizon), the steering between Alice and Anti-Rob is bidirectional, whereas when $R_{0}$ exceeds a critical value, the steering becomes unidirectional.

Finally, by further tracing the pattern in Alice's local region, the reduced density matrix between Rob and Anti-Rob can be obtained as
\begin{equation}
	\hat{\rho}_{R\bar{R}} = 
	\begin{pmatrix}
		\frac{1}{2}\cos^{2}\alpha & 0 & 0 & \frac{1}{2}\sin\alpha\cos\alpha \\
		0 & 0 & 0 & 0 \\
		0 & 0 & \frac{1}{2} & 0 \\
		\frac{1}{2}\sin\alpha\cos\alpha & 0 & 0 & \frac{1}{2}\sin^{2}\alpha
	\end{pmatrix}.
	\label{eq:rho_RR}
\end{equation}

 We can also examine the quantum steering relations between Rob and Anti-Rob, as shown in Fig.\ref{fig2} (b). 
At sufficiently low values of $\omega$, the horizontal axis represents the steering from Anti-Rob to Rob $S^{\bar{R} \to R}$, whereas the vertical axis corresponds to the steering from Rob to Anti-Rob $S^{R \to \bar{R}}$.
Interestingly, it is found that the quantum steering between Rob and Anti-Rob is unidirectional. 
Specifically, for fixed $R_{0}$, the steering measure $S^{R \to \bar{R}}$ decreases with increasing $\ell$. 
However, $S^{\bar{R} \to R}$ remains zero as $\ell$ increases, indicating the absence of steering from Anti-Rob to Rob. 
As observers move farther from the event horizon of the black hole, the influence of increasing $\ell$ on the steering from Rob to Anti-Rob becomes more pronounced. 
This demonstrates that the Lorentz-violating parameter  has a significant impact on quantum steering both inside and outside the black hole’s event horizon.
Furthermore, this result is crucial for our subsequent analysis of Lorentz-violation effects in quantum field theory in curved spacetime.

It is shown that the steering from one particle to another are always unequal. Therefore,  we define the mode pairs $A$ and $R$, $A$ and $\bar{R}$, $R$ and $\bar{R}$, for which the corresponding directional asymmetries are defined by
\begin{align}
	\Delta S_{AR}   &= \left| S^{A \rightarrow R} - S^{R \rightarrow A} \right|, \\
	\Delta S_{A\bar{R}} &= \left| S^{A \rightarrow \bar{R}} - S^{\bar{R} \rightarrow A} \right|, \\
	\Delta S_{R\bar{R}} &= \left| S^{R \rightarrow \bar{R}} - S^{\bar{R} \rightarrow R} \right|.
\end{align}

\begin{figure}[h]
	\centering
	\includegraphics[width=0.7\linewidth]{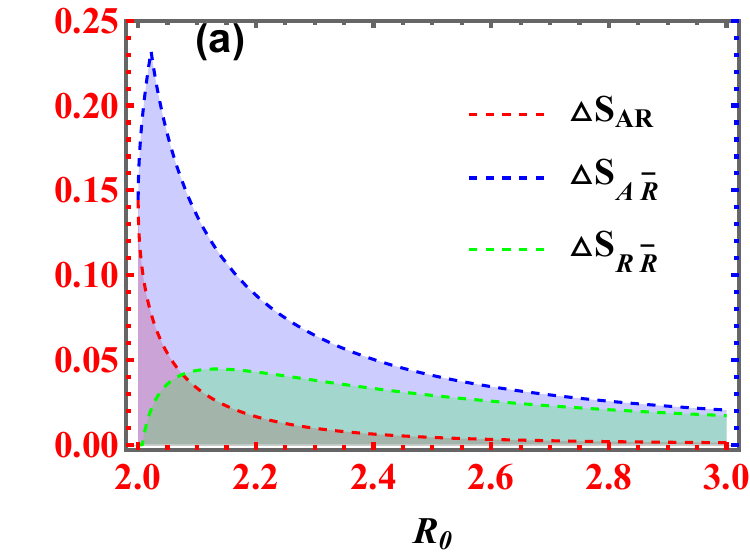}
	\caption{The values of $\Delta S_{AR}$, $\Delta S_{A\overline{R}}$, and $\Delta S_{R\overline{R}}$ as a function of $R_{0}$. Here, $\omega=0.1$ and $\ell=0.1$.}
	\label{fig4}
\end{figure}

In Fig. \ref{fig4}, we illustrate the dependence of steering asymmetry on $R_{0}$. 
It is found that within a certain range of $R_{0}$ values, the three types of steering asymmetry vary with increasing $R_{0}$. Consequently, quantum steering between any two states remains asymmetric at a fixed distance from the black hole's event horizon. Among these,  $\Delta S_{AR}$ decreases monotonically with increasing $R_{0}$ and eventually vanishes far from the horizon. In contrast, $\Delta S_{A\overline{R}}$ first reaches a maximum and then decreases to a constant value, while $\Delta S_{R\overline{R}}$  displays a similar nonmonotonic behavior. As  $R_{0}$ increases, Rob's intrinsic acceleration decreases, and the equivalent gravitational pull from the black hole's surface diminishes, resulting in nearly symmetric steering between particles. This further indicates that, in addition to the Lorentz-violating parameter, the gravitational acceleration at the black hole surface plays a crucial role in determining fermionic quantum steering.

In a word, the quantum steering between two subsystems depends sensitively on variations in the acceleration parameter at the event horizon. 
This behavior indicates that the underlying Fermi–Dirac distribution plays a crucial role in determining the system’s quantum steering properties. 
Moreover, for the tripartite configuration involving Alice, Rob, and Anti-Rob, we find that, for fixed acceleration, the introduction of a Lorentz-violating parameter induces directional asymmetry in quantum steering.  In particular, quantum steering is confined to a narrow region near the event horizon, and Lorentz violation further restricts the extent of this region. 
Therefore, these results further confirm that Lorentz violation has a nontrivial influence on quantum steering.

\section{Bell's Nonlocality and the CHSH Inequality} \label{sec5}
To investigate the relationship between quantum steering and Bell nonlocality  in the curved spacetime, we employ the CHSH inequality to examine the behavior of Bell nonlocality. 
To obtain CHSH inequality, one need to   introduce the CHSH operator \cite{Sun:2017agp}
\begin{equation}
	\mathcal{B} = \mathbf{a} \cdot \boldsymbol{\sigma} \otimes \left( \mathbf{b} + \mathbf{b}^{\prime} \right) \cdot \boldsymbol{\sigma} + \mathbf{a}^{\prime} \cdot \boldsymbol{\sigma} \otimes \left( \mathbf{b} - \mathbf{b}^{\prime} \right) \cdot \boldsymbol{\sigma}
	\label{eq:bell_operator},
\end{equation}
where
$\mathbf{a}=(a_{x},a_{y},a_{z}),\mathbf{a},\mathbf{a}^{\prime},\mathbf{b},\mathbf{b}^{\prime}$  denote the unit vectors to be observed in two-body state, while $\boldsymbol{\sigma}=(\sigma_{x},\sigma_{y},\sigma_{z})$ represents the vector within the Pauli operators.
The product of these two vectors can be written as 
\begin{equation}
	\mathbf{a} \cdot \boldsymbol{\sigma} = \sum_{i} a_{i} \sigma_{i}.
	\label{eq:dot_product_summation}
\end{equation}
The validity of the CHSH inequality can be examined  by computing the reduced density matrix of an arbitrary two-body state \cite{Sun:2017agp}, thereby characterizing the correlations between the two-body state.
According to equation (\ref{eq:bell_operator}), the Bell inequality can be written as
\begin{equation}
	B(\rho) = |\operatorname{Tr}(\rho B)| \leq 2.
	\label{eq:bell_inequality}
\end{equation}
If $B(\rho) \geq 2$, the two-body state exhibits Bell nonlocality.
The maximal Bell function is given by
\begin{equation}
	B(\rho) = 2\sqrt{\max_{i<j}(K_i + K_j)},
	\label{eq:B_rho_formula}
\end{equation}
where $K_{i}$ and $K_{j}$ denote the eigenvalues of a symmetric matrix. 
Then ew define $K(\rho)=T_{\rho}^{T}T_{\rho}$, where $T=(t_{ij})$ is the correlation matrix with elements $t_{ij} = \operatorname{Tr}[\rho \sigma_{\mathrm{i}} \sigma_{\mathrm{j}}]$.
In the standard two-qubit basis, the eigenvalues are expressed as
\begin{equation}
	K_{1} = 4\left(\left|\rho_{14}\right| + \left|\rho_{23}\right|\right)^{2,}
	\label{eq:K1}
\end{equation}
\begin{equation}
	K_{2} = 4\left(\left|\rho_{14}\right| - \left|\rho_{23}\right|\right)^{2},
	\label{eq:K2}
\end{equation}
\begin{equation}
	K_{3} = \left(\left|\rho_{11}\right| - \left|\rho_{22}\right| - \left|\rho_{33}\right| + \left|\rho_{44}\right|\right)^{2}.
	\label{eq:K3}
\end{equation}
The maximal Bell function is therefore given by
\begin{equation}
	B\left(\rho_{x}\right)=\max \left\{ B_{1}, B_{2} \right\},
	\label{B_rho_equation}
\end{equation}
where $B_1 = 2\sqrt{K_1 + K_2}$ and $B_2 = 2\sqrt{K_1 + K_3}$ \cite{Shadbolt:2011kyq,Sun:2017agp}. 
This confirms that the maximal violation of the CHSH inequality for a two-qubit state is $2\sqrt{2}$. 
Based on these parameter relations, the Bell locality properties of the two-particle state can be examined. 
According to equation (\ref{eq:B_rho_formula}), the maximal Bell functions for the bipartite modes $A$ and $R$, $A$ and $\bar{R}$, $R$ and $\bar{R}$  are obtained as 
\begin{equation}
	\begin{aligned}
		B(\rho_{AR}) &= \max\bigg\{ 2\sqrt{2}\,|\cos R|, \\
		&\quad 2\sqrt{|\cos R| \left[ 2 + \left( \frac{1}{2} + \frac{\cos^2 R}{2} - \frac{\sin^2 R}{2} \right)^2 \right] } \bigg\},
	\end{aligned}
\end{equation}
\vspace{-\baselineskip}
\begin{equation}
	\begin{aligned}
		B(\rho_{A\overline{R}}) &= \max\bigg\{ 2\sqrt{2}\,|\sin R|, \\
		&\quad 2\sqrt{|\sin R| \left[ 2 + \left( -\frac{1}{2} + \frac{\cos^2 R}{2} - \frac{\sin^2 R}{2} \right)^2 \right] } \bigg\},
	\end{aligned}
\end{equation}
\vspace{-\baselineskip} and
\begin{equation}
	\begin{aligned}
		B(\rho_{R\overline{R}}) &= \max\bigg\{ \sqrt{2}\,|\sin(2R)|, \\
		&\quad 2\sqrt{ \frac{1}{4}\left( -1 + |\cos R|^2 + |\sin R|^2 \right)^2 + |\cos R \sin R|^2 } \bigg\}.
	\end{aligned}
\end{equation}

\begin{figure}[h]
	\centering
	\includegraphics[width=0.7\linewidth]{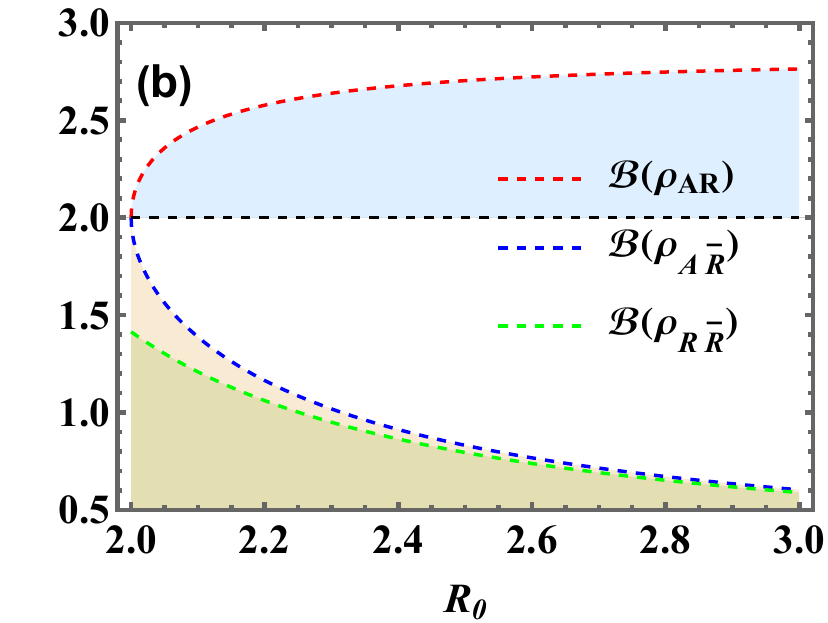}
	\caption{The values of $B(\rho_{AR})$, $B(\rho_{A\overline{R}})$, and $B(\rho_{R\overline{R}})$  as a function of $R_{0}$. Here, $\omega=0.1$ and $\ell=0.1$.}
	\label{fig5}
\end{figure}

As shown in Fig. \ref{fig5}, at sufficiently low values of $\omega$ and for a fixed $\ell$, the vertical axis represents the maximum Bell signal as a function of $R_{0}$.
It can be observed that within a certain range of $R_{0}$ values, as $R_{0}$ increases, $B(\rho_{AR})$ monotonically increases and consistently violates the Bell inequality. 
This demonstrates Bell nonlocality under the entanglement formed between Alice and Rob, indicating that fermionic Bell nonlocality is dependent on $R_{0}$. In addition, the $B(\rho_{A\overline{R}})$  monotonically decreases with increasing $R_{0}$ and consistently satisfies the Bell inequality, indicating that no Bell nonlocality is exhibited between Alice and Anti-Rob. Similarly, $B(\rho_{R\overline{R}})$ monotonically decreases with increasing $R_{0}$ and consistently satisfies the Bell inequality, indicating that no Bell nonlocality is exhibited between Rob and Anti-Rob.

By comparing the relationship between any two particle states, we find that within the physically accessible region, the stronger the Bell nonlocality exhibited, the greater the increase in $R_{0}$. This demonstrates that Bell's inequality can transcend the locality previously assumed, indicating that quantum mechanics can break through traditional perspectives to achieve nonlocal correlations.

\section{CONCLUSIONS AND OUTLOOKS} \label{sec6}
This study systematically investigates the influence of the Lorentz-violating parameter on nonlocal quantum correlations of Dirac fields in a static Einstein-Bumblebee black hole spacetime. By tracing over the modes of Alice, Rob, and Anti-Rob, we have analyzed quantum steering and Bell nonlocality in this Lorentz-violating gravitational background. Our results demonstrate that the Lorentz-violating parameter induces a pronounced steering asymmetry, which exhibits distinct behavioral patterns based on the observer's position relative to the event horizon. Specifically, in the physically accessible region, steering asymmetry initially increases with distance from the horizon, reaching a maximum value before gradually decreasing at larger separations. Furthermore, our analysis of Bell nonlocality reveals a consistent violation of Bell inequalities in the accessible region, with the degree of nonlocality strengthening as the distance from the event horizon increases. 
The  findings is the present work not only deepen our understanding of quantum correlations in curved spacetime with Lorentz violation but also provide theoretical foundations for selecting quantum resources in relativistic quantum information processing tasks. 

Future research could extend our approach to rotating black hole spacetimes or dynamic gravitational backgrounds to examine how additional parameters influence quantum correlations.  Investigating other quantum resources and applying this framework to additional modified gravity theories would further elucidate the interplay between quantum information and fundamental symmetry violations. Finally, connecting these theoretical predictions with potential observational signatures in analogue gravity experiments may open pathways for empirical validation of quantum gravitational effects.

\begin{acknowledgments}
	This work was supported by the National Natural Science Foundation of China (Grants No. 12475051, No. 12374408,  and No. 12547147); the science and technology innovation Program of Hunan Province under grant No. 2024RC1050;  the Natural Science Fund of Hunan Province under Grant No. 2026JJ20019; and the China Postdoctoral Science Foundation (Grant No. 2025M783393).
\end{acknowledgments}

%

\end{document}